\documentclass[runningheads]{llncs}
\usepackage{graphicx}
\usepackage{import}
\usepackage{algorithm}
\usepackage{amssymb}
\usepackage{esvect}
\usepackage{multicol}
\usepackage{subfigure} 
\usepackage{ulem}
\usepackage{algorithmic}
\usepackage[cmex10]{amsmath}
\usepackage{eqparbox}

\begin{document}
\title{Mobile Crowdsourced Sensors Selection for Journey Services}

\author{Ahmed Ben Said\inst{1} \and
Abdelkarim Erradi\inst{1} \and
Azadeh Gharia Neiat\inst{2} \and 
Athman Bouguettaya\inst{2}}
\authorrunning{A. Ben Said et al.}

\institute{Department of Computer Science and Engineering, College of Engineering, 
\newline Qatar University, Doha, Qatar
\email{\{abensaid,erradi\}@qu.edu.qa}\\
\and
School of Information Technologies, University of Sydney, Australia\\
\email{\{azadeh.gharineiat,athman.bouguettaya\}@sydney.edu.au}}
\maketitle              
\begin{abstract}
We propose a  mobile crowdsourced sensors selection approach to improve the journey planning service especially in areas where no wireless or vehicular sensors are available. We develop a location estimation model of journey services based on an unsupervised learning model to select and cluster the right mobile crowdsourced sensors that are accurately mapped to the right journey service. In our model, the mobile crowdsourced sensors trajectories are clustered based on common features such as speed and direction. Experimental results demonstrate that the proposed framework is efficient in selecting the right crowdsourced sensors. 

\end{abstract}
\section{Introduction}
\begin{figure*}[t!]
	\centering
	\includegraphics[scale=.6]{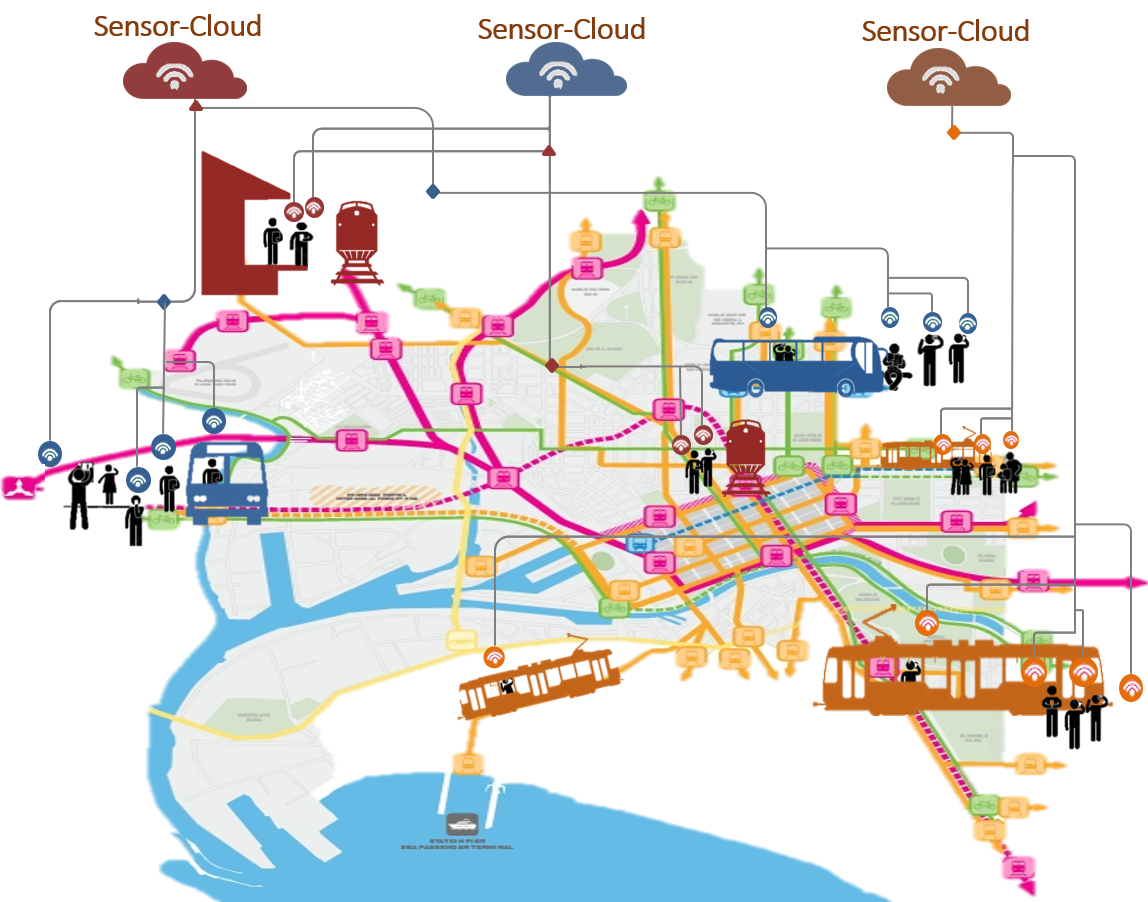}
	\caption{Overview of public transport scenario powered by MCS}
	\label{scenario}
\end{figure*}

With the increasing use of mobile devices such as smartphones and wearable devices  mobile crowdsourced sensing is emerging as a new sensing paradigm for obtaining required sensor data and services by soliciting contribution from the crowd. Storing, processing and managing continuous streams of crowdsensed data pose key challenges. The cloud offers a new paradigm, called sensor cloud \cite{wsn,azadeh1} to efficiently handle these challenges. 
\newline
In our previous works \cite{azadeh3} \cite{azadeh2} \cite{azadeh_failure_proof}, we explore a new area in spatio-temporal travel planning by abstracting the problem using the service paradigm. We assume that we have a map consisting of spatial routes which in turn consist of segments. Each sensor cloud segment service is served by a journey vehicle (e.g., buses, trams or trains) and has a number of attributes and associated quality of service. Th functional attributes of a line segment service include  GPS coordinates of the source and destination points and the mode of transportation (e.g. train service). Quality of service parameters include times of arrival and departure, accuracy, cost and so on. Therefore, a journey would consist of composing a set of line segment services on the map according to a set of functional and non-functional requirements. In \cite{azadeh3} \cite{azadeh2} \cite{azadeh_failure_proof}, we made the assumption that the sensor cloud services are modeled by fixed sensors, i.e. sensors embedded on the journey vehicle e.g. bus, tram. 
\newline 
In this paper, we consider the scenario illustrated in Fig. \ref{scenario} in which the crowd is the source of information. Indeed, we solely rely on commuters providing real-time geolocation data collected through their mobile devices (e.g. smartphone) instead of fixed sensors, called Mobile Crowdsourced Sensors (MCS). These MCS are abstracted on the cloud and can be used by the journey planning service to serve commuters' requests to find optimal journey plans. This is an interesting development since MCS represent an alternate source of information in absence of sensory infrastructure and eliminate the need to deploy costly sensory equipment. Commuters representing the source of crowdsourced sensors are willing to participate if they are convinced and well-incentivized, i.e. they are provided with a reward either as a service compensation or money \cite{incentive}. For example, participants can benefit from enhanced journey planning and real-time transport network update using the collected data. In addition, they can get credit compensation according to the level of participation. In this scenario, we suppose that commuters are well-incentivized to participate in sharing their real-time locations.
\newline
The key challenge of leveraging MCS is to identify and access the “right” crowdsourced sensors that are applicable to a particular journey planning request. This is important as MCS will move from one mode of transportation to another, i.e., they would serve different sets of sensor cloud services at different locations and times. As a result, there is a need to develop an efficient selection technique to accurately identify MCS that will be mapped to the right sensor cloud service. 
\newline
This paper focuses on developing a MCS selection technique to enable an optimal journey planning powered by MCS. Specifically, we propose an unsupervised learning approach to select and cluster the right mobile crowdsourced sensors based on common patterns in their trajectories. \newline
The main contributions of this paper are: (1) A new formal model of mobile crowdsourced sensors allowing access to the right MCS in space and time to enable identifying and tracking the location of journey vehicles. (2) A new unsupervised learning approach for clustering MCS. (3) Novel quality measures based on moving characteristic of MCS to assess the homogeneity of members of identified crowdsourced sensor clusters.
\newline
The rest of the paper is organized as follows: in section II, we review relevant related works. In section III, we present the problem formulation and the proposed model. In Section IV, we discuss the details of the proposed crowdsourced sensors selection and clustering algorithm that allows the identification of journey vehicles. In section V, we present and analyze the results of the experimental evaluation. Finally, conclusions and future work are presented in the last section. 
\section{Related Work}
Several research proposals have focused on mobile crowdsourcing for journey planning service. Yu et al. \cite{recommender} proposed a MCS-based travel package recommendation system. A profile is constructed for each user to leverage spatio-temporal features of check-in in points of interests (POI) which are hierarchically classified. Each POI is characterized by its periodic popularity. These information are then used in real-time to recommend personalized travel packages while taking into account user preferences, POI characteristics, and spatio-temporal constraints such as travel time and initial location. Chen et al. \cite{tripplanner} used MCS to build  the pattern map of the metro line, which can then be used for localization. The system consists of two phases: in the first one,  patterns from user traces are extracted, and mined to identify the ones which are linked to specific tunnels. This allows to construct the graph of the metro line. In the second phase, the pattern map is made available on the cloud for users to download. When user travels using metro line, barometric pressure and magnetic fields data are logged along with stop and running events. Therefore, the train and user locations are known. Shin et al. \cite{citying} proposed a MCS-based approach for classification of transport mode. Authors collected information including date and time, x-, y-, and z-acceleration values, latitude and longitude. By analyzing these records, the walking pattern is characterized and used to segment the overall activities. To determine the travel mode of a vehicle-ride activity, the acceleration profile for each mode is estimated and used to classify particular acceleration behavior into one of the modes. In \cite{trafficinfo}, authors proposed TrafficInfo, a participatory sensing based live public transport information service. Instead of relying costly sensing infrastructure, the proposed service relies on contribution from the crowd to visualize the actual position of the journey vehicle.\newline
The aforementioned works consider the mapping between MCS and journey vehicle (tram, metro, etc ...) as a prerequisite or assume that the crowd are fully cooperative and handle this mapping even though they can move from one mode of transportation to another or share erroneous information. However, this assumption does not always hold. Indeed, such task requires an effective incentive mechanism to motivate the participants. Furthermore, the crowd can be  indifferent to handle this task or introduce erroneous information. Therefore it is essential to develop techniques for automatic selection of the right MCS that enable the estimation of the journey vehicle location in real time. 
\newline
The availability of data enjoying spatio-temporal properties has elicited new data analysis paradigm to explore and extract new spatio-temporal patterns. 
Spatiotemporal clustering is the process of grouping data objects based on space and time relationships. Spatiotemporal clustering methods determine to which cluster a given object belongs based on different features such as the speed, the direction and the similarity in the trajectory origin and destination. 
Since trajectory is a sequence of time-stamped location points of a moving object through space, grouping moving trajectories is complex due to their continuous movement. Thus, more efforts are needed to discover the interaction and change in the spatiotemporal trajectory movements in order to achieve more accurate partitioning \cite{clustering1}. Recently, researchers are proposing modifications of existing clustering algorithms to make them more suitable for spatiotemporal data. Birant et al. \cite{clustering2} proposed a spatio temporal algorithm called ST-DBSCAN, an extension of the well known DBSCAN algorithm to the spatiotemporal domain. Avni et al. \cite{clustering3} also extended the Ordering Points to Identify the Clustering Structure (OPTIC) algorithm to cluster spatiotemporal data for taxi recommendation system. On the other hand, spatiotemporal pattern mining focuses on discovering hidden movement patterns from the trajectories of moving objects. Multiple methods were proposed to mine several types of movement patterns for groups of objects that move together in a near space and time. These patterns include periodic or repetitive pattern that concerns regular movement e.g. bird migration \cite{clustering6,clustering7}, flock \cite{flock}, convoy \cite{convoy}, swarm \cite{clustering6}, leadership \cite{leadership} and chasing \cite{chasing}.
\newline
The evaluation of spatiotemporal clustering approaches remains an open and challenging issue. While the traditional clustering approaches require computation in single Euclidean space, the spatio-temporal clustering approaches need computation in multiple spaces \cite{clustering4}. In addition, grouping spatio-temporal data is affected by the large data size which leads to a trade-off between  accurate clustering results and computational cost \cite{clustering5}. Clustering is also affected by noise and outliers. Additionally, the presence of clusters of different shapes, e.g. ellipsoid, and of unbalanced sizes may result in the inaccurate data partition. Indeed, some clustering algorithms, e.g. K-means, form clusters with a circular shape which leads to misleading results.  
\section{System model}
\begin{figure}[ht]
	\centering
	\includegraphics[scale=.12]{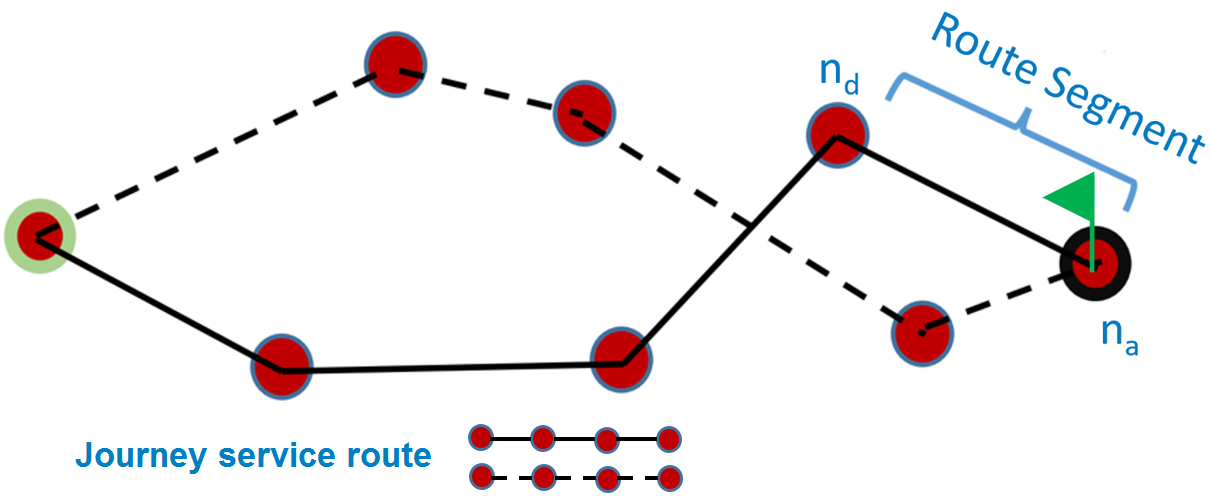}
	\caption{A journey network consists of journey service routes which in turn consist of route segments}
	\label{segment}
\end{figure}

Our objective is to identify journey vehicles and track their location by relying solely on crowdsourced sensors. Therefore, it is important to select the right subset of crowdsourced sensors. In this context, a group of MCS associated to a journey vehicle are very likely to have similar spatiotemporal features. Our strategy consists of grouping or clustering the set of MCS that have common patterns.\newline
In the following, we first define preliminary concepts and then present the problem statement. In the remainder, we refer to sensor cloud service as a journey service. A journey network is a spatial representation of journey services (i.e., bus, train etc) available in a given area. It consists of journey service routes composed of route segments (see Fig. \ref{segment}). We formally define a route segment and journey service as follows:

\textbf{Definition 1: Route segment.} A route segment ($rs_i$) connects two nodes representing the source and destination points. It is identified by a tuple $<n_d, n_a, dist, speed, att>$ where:

\begin{itemize}
	\item $n_d$: GPS coordinates of the departure node.
	\item $n_a$: GPS coordinates of the arrival node.
	\item $dist$: Distance between $n_d$ and $n_a$.
	\item $speed$: Average travel speed along the route segment.
	\item $att$: Average travel time duration to traverse the route segment.
\end{itemize}

\textbf{Definition 2: Journey Service.} A Journey Service (JS) is modeled as a composition of route segments. JS is described by a tuple $<JS_{id}, RS, schedule>$ where:\newline
- $JS_{id}$: A unique identifier of the service. We consider each inbound or outbound direction as a separate service (e.g., Bus-30-Manhattan-JFK-Express is a service). \newline
- $RS$: A list of route segments \{$rs_1$, $rs_2$, ... $rs_n$\} comprise the service route.\newline
- $schedule$: Scheduled JS trips per day. It can also be represented by an average headway which is defined as the time difference between any two successive vehicles. \newline
A journey service could be served by one or more Journey Vehicles (JV) such as buses. 

\textbf{Definition 3: Journey Vehicle.} A JV is identified by a tuple $<v_s, dt, rs, loc, t_d, t_a>$ where:
\begin{itemize}
	\item $v_s$: Journey service Id (e.g., bus line 100) whose  journey vehicle is currently serving.
	\item $dt$: Departure time from the start of the journey.
	\item $rs$: Current route segment being traversed by the vehicle.
	\item $loc$: Current vehicle location.
	\item $t_d$: Departure time for the current route segment.
	\item $t_a$: Estimated arrival time to the next stop which is the arrival node of the current route segment.
\end{itemize}
Crowdsourced sensors are constantly sending their location information to a cloud-hosted journey service. These spatio-temporal records can be modeled as MCS trajectory. Given a set of crowdsourced sensors represented by their trajectories $\mathcal{S} = \{ Tr_1, Tr_2,... , Tr_N \}$, where $N$ is the number of trajectories, the proposed algorithm discovers clusters of crowdsourced sensors $\mathcal{C} = \{ C_1, C_2,... , C_M \}$, where $M$ is the number of cluster centers.

\begin{figure}[ht]
	\centering
	\includegraphics[scale=.27]{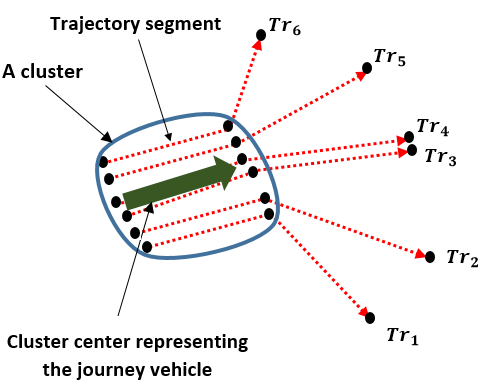}
	\caption{An example illustrating a journey vehicle identification by clustering the crowdsourced sensor trajectories}
	\label{cs}
\end{figure}

\textbf{Definition 4: Crowdsourced sensor trajectory.} A crowdsourced sensor trajectory $Tr_i$ is a set of sequential timestamped geolocations:\newline $Tr_i = [(p_1, t_1) ,(p_2, t_2)... (p_L, t_L)]$. A geolocation $p_i$ is a pair of latitude and longitude sent by the sensor at time $t_i$. $L$ is the trajectory length. It can be different from one trajectory to another. We assume that trajectories are defined for the same time intervals. A trajectory segment is a pair of consecutive timestamped geolocations: $ts=[(p_j,t_j),(p_k,t_k)],{t_j<t_k}.$ A trajectory of length $L$ is composed of $L-1$ trajectory segments. It is also characterized by its associated direction and speed. A cluster of crowdsourced sensors is a group of sub-trajectories as illustrated in Fig. \ref{cs}.

\vspace{- 0.5 cm}
A cluster center is an imaginary trajectory segment with specific characteristics i.e. start and end point as well as start and end time. This particular sub-trajectory is the representation of the journey vehicle.

\begin{figure}[t]
	\centering
	\includegraphics[scale=.25]{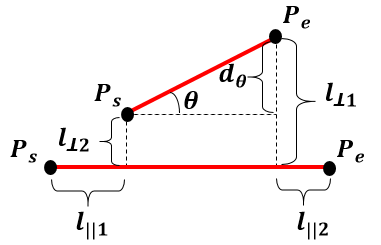}
	\caption{The distance measure between two trajectory segments}
	\label{distance}
\end{figure}

\textbf{Definition 5: Distance Function.} A clustering algorithm, whether density based, partitional or hierarchical, is formulated using a distance measure. To take the particularity of our trajectory data structure into account, we propose to use the distance measure proposed by Lee et al. \cite{distance} illustrated in Fig. \ref{distance}. Specifically, the distance between trajectory segments $ts_1$ and $ts_2$ is a linear combination of three distances: perpendicular distance $d_\perp$, parallel distance $d_{||}$ and angle distance $d_{\theta}$:
\begin{equation}
d(ts_1, ts_2)=d_\perp(ts_1, ts_2) + d_{||}(ts_1, ts_2) + d_\theta(ts_1, ts_2)
\end{equation}
where:
\begin{equation}
d_\perp(ts_1, ts_2) = \frac{l_{\perp1}^2+l_{\perp2}^2}{l_{\perp1}+l_{\perp2}}
\end{equation}
\begin{equation}
d_{||}(ts_1, ts_2) = \min(l_{||1},l_{||2})
\end{equation}   
\begin{equation}
d_\theta(ts_1, ts_2) = length(ts_2)\times sin(\theta)
\end{equation} 
For more accurate distance measure between two geolocations, the Haversine distance or Vincenty distance \cite{geodistance} can be used. However, for simplicity, we use the classic Euclidean distance to calculate $l_{||1}$, $l_{||2}$, $l_{\perp1}$ and $l_{\perp2}$ which is suitable for the small area of study used for evaluation.

\textbf{Definition 6: $\epsilon$-Neighborhood.} The neighborhood of a trajectory segment $ts_i$ with respect to $\epsilon > 0$, denoted $N_{\epsilon}(ts_i)$, is a subset of trajectory segments and defined as:
\begin{equation}
N_{\epsilon}(ts_i) = \{ ts_j \;|\; d(ts_i,ts_j) \leq \epsilon\}
\end{equation}
It is the set of trajectory segments whose distance to $ts_i$ is less than a threshold $\epsilon$. 

\textbf{Definition 7: Core Trajectory Segment.}
A trajectory segment $ts_i$ is a core trajectory segment defined with respect to $\epsilon$ and $MinS >0$ iff:
\begin{equation}
|N_{\epsilon}(ts_i)| \geq MinS
\end{equation}
Where $|N_{\epsilon}(ts_i)|$ is the cardinality of $N_{\epsilon}(ts_i)$.
A core trajectory segment highlights the presence of dense region. $MinS$ is a minimum number of neighbor trajectory segments required to form a dense region around $ts_i$.

\begin{figure}[ht]
	\centering
	\includegraphics[scale=.2]{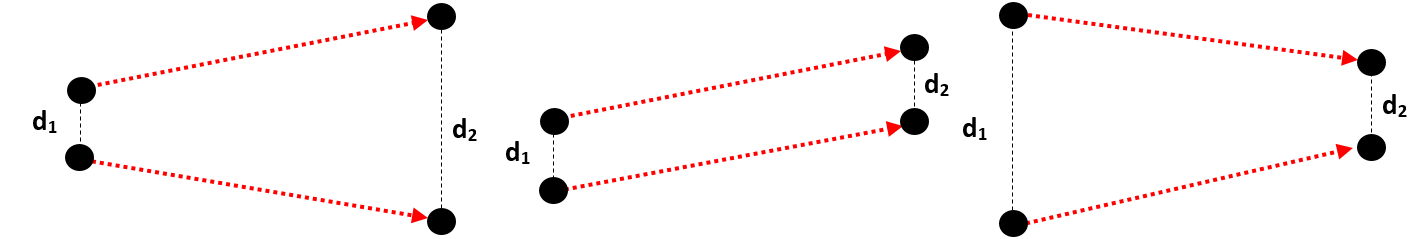}
	\caption{Three following scenarios}
	\label{following}
\end{figure}

\textbf{Definition 8: Following Degree $\textbf{(FD)}$.}
The following degree between two trajectory segments is established given the three possibilities illustrated in Fig. \ref{following}. It takes into account whether the trajectory segments are converging, diverging or parallel. It is defined as follows:
\begin{equation}
FD=\begin{cases}
\;\;1, & if \;\; d_1=d_2\\
\;-1, & if \;\; d_1 < d_2\\
\;\;\frac{d_2}{d_1}, & if \;\; d_1 > d_2\\
\end{cases}
\end{equation}
If the two trajectory segments originate from the same geolocation i.e. $d_1=0$, the trajectory segments are diverging and $FD=-1$. For the special case where $d_1=d_2=0$, $FD$ is equal to 1.

\textbf{Definition 9: Trajectory Segment Direction $\textbf{(DR)}$.}
The trajectory segment direction is defined as the  counterclockwise angle  of the trajectory segment with respect to the reference line of Equator. The direction can be also derived from the accelerometer and the geomagnetic field sensor embedded on the smartphone.

\textbf{Definition 10: Trajectory Segment Speed \textbf{$\textbf{(SP)}$}.}
The trajectory segment is characterized by its speed. It is the distance between the departure node $n_d$ and the arrival node $n_a$ over the difference in time: $t_k-t_j$.

\section{Mobile crowdsourced sensors selection algorithm}
In this section, we propose our spatio-temporal crowdsourced sensors clustering algorithm for crowdsourced sensors selection. We first present the details of the algorithm. Then, we define the homogeneity score which is used to form  clusters.

\subsection{Spatio-temporal crowdsourced sensors clustering algorithm}

In classic clustering task, a high performance algorithm achieves a partition of objects where members of each cluster are as homogeneous as possible with respect to certain criteria. For example, a cluster should be as dense as possible. The density can be quantified using a discrepancy measure such as the variance. \newline
Our strategy for trajectory segment clustering originates from the following intuitive idea: MCS contributing in identifying a journey vehicle associated to a journey service should (1) be as dense as possible and (2) share common spatio-temporal patterns such as speed, following degree and direction. Consequently, our algorithm seeks to identify dense regions with respect to predefined parameters: $\epsilon$ and $MinS$. These regions indicate the presence of potential clusters. However, it is important to identify clusters with the highest homogeneity among each member. Therefore, we define a homogeneity score. It captures the spatio-temporal dynamism of trajectory segments such as speed, direction and following degree and then identifies the cluster with the highest homogeneity. \newline
Our algorithm, detailed in \ref{algo1},  first identifies the list of core trajectory segments (line 3-6) at each timestamp. Then, it considers every core trajectory segment $ts_i$ and its $\epsilon$-neighborhood as a potential cluster. It seeks also to form clusters as homogeneous as possible with respect to a particular score called the Homogeneity Score ($HS$). A potential cluster is added to the set of clusters if it fulfills one of the following requirements:
\begin{itemize}
	\item Its associated core trajectory segment has no other core trajectory segment in the list of its $\epsilon$-neighborhood (line 10-12)
	\item Its associated core trajectory segment has the lowest $HS$ among the neighbor core trajectory segments (line 14-17).
\end{itemize}
After a cluster of trajectory segments is formed, its cluster center can be used to identify the associated journey service. In this regard, we consider the  vectorized version of the trajectory segments $\vv{ts_1}$, $\vv{ts_2}$, ..., $\vv{ts_N}$ where:
\begin{align}
\vv{ts_i} &= \begin{pmatrix}
x_{j+1} - x_{j} \\
y_{j+1} - y_{j} \\
t_{j+1} - t_{j}
\end{pmatrix}
\end{align}
The average trajectory segment of $N$ trajectory segments is defined as:
\begin{equation}
\vv{cc} = \frac{\vv{ts_1}+\vv{ts_2}+ ... +\vv{ts_N}}{N}
\end{equation}

The journey vehicle is identified by the trajectory segment associated with $\vv{cc}$. Therefore, the spatio-temporal properties of the cluster centers correspond to the journey vehicle properties. By gradually capturing the set of cluster center and therefore the journey vehicle, we continuously update the journey service and route segment attributes such as the speed and arrival time.
\begin{algorithm}
	\caption{Crowdsourced Sensors Clustering}
	\label{algo1}
	\textbf{Input:} Trajectory set $\mathcal{S}$, $\epsilon$, $MinS$\\
	\textbf{Output:} Identified journey vehicles\\
	\begin{algorithmic}[1]
		\STATE List$_{CC} \; \leftarrow \; \varnothing \; \; \;$ \textit{ \# list of identified vehicles}  \\
		\COMMENT {\textit{\# Identify core trajectory segments}}\\ 
		\FOR{every time slot $\Delta t$ } 
		\FOR {$ts_i \in \mathcal{S}$ } 
		\IF {  $ts_i$ is a core trajectory segment (Definitions \textbf{5} 					and \textbf{6})}
		\STATE {Insert $ts_i$ in $List_{CS}$}
		\STATE {Calculate $HS(ts_i)$ using (\textbf{\ref{hs}})}
		\ENDIF
		\ENDFOR
		\ENDFOR \\
		\COMMENT{\# \textit{Cluster trajectory segments}}\\
		\FOR {every $ts_i \in List_{CS} $} 
		\IF {$\forall \;ts_j\;\in \;N_{\epsilon}(ts_i);\;ts_j\;\not\in 						List_{CS}$}
		\STATE {Calculate the cluster center $cc$}
		\STATE{Insert \big($ts_j$, $N_{\epsilon}(ts_j)$, $cc$\big) in 						$List_{CC}$}
		\ELSE
		\FOR{ \textbf{each} $ts_j \in \big\{N_{\epsilon}(ts_i)\:\cup\:{ts_i}\big\}\: \& \: ts_j \in List_{CS}$}
		\STATE {Identify $ts_j$ with the lowest $HS$} 
		\STATE{Calculate the cluster center $cc$ }
		\STATE{Insert \big($ts_j$, $N_{\epsilon}(ts_j)$, $cc$\big) 							in $List_{CC}$}
		\ENDFOR
		\ENDIF
		\ENDFOR\\    
	\end{algorithmic}
\end{algorithm}
\subsection{The Homogeneity Score}
The Homogeneity Score $HS$ is of paramount importance since it identifies the clusters and therefore the journey vehicles. $HS$ is a combination of three scores: the following, the speed and the direction scores. 
It captures the spatio-temporal properties of MCS. Indeed, members of a cluster are supposed to follow each other with relatively the same speed and direction. 
These scores are defined as follows:\newline

\noindent\textbf{Definition 11: Following Score \textbf{(FS)}.}
The following score $FS$ of a core trajectory segment $ts_i$ is defined as:
\begin{equation}
FS = |N_{\epsilon}(ts_i)| - \sum_{j=1,j\\\neq i}^{|N_{\epsilon}(ts_i)|}FD(ts_i,ts_j)
\end{equation}
It evaluates how different the following score of the core trajectory segment to the average following score of the cluster. A low $FS$ score indicates better homogeneity in terms of following.\newline

\noindent\textbf{Definition 12: Speed Score \textbf{(SS)}.} It measures the homogeneity of the cluster in terms of speed. Indeed, a group of crowdsourced sensors should move with the homogeneous velocity. Given a potential cluster defined with respect to core trajectory segment $ts_i$, $SS$ is expressed as:
\begin{equation}
SS=\frac{\big|\;SP(ts_i) - \frac{1}{|N_{\epsilon}(ts_i)|}\sum\limits_{j}^{|N_{\epsilon}(ts_i)|}SP(ts_j)\;\big|}{\max\limits_{j} \big\{SP(ts_j)\big\} - \min\limits_{j} \big\{SP(ts_j)\big\} }
\end{equation}
$SS$ evaluates the difference between the core trajectory segment speed and the average speed of the clusters. A low $SS$ score indicates better homogeneity in term of speed.\newline

\noindent\textbf{Definition 13: Direction Score \textbf{(DS)}.} It measures the homogeneity of the cluster in terms of direction. A group of crowdsourced sensors should move in the same direction. Given a potential cluster defined with respect to core trajectory segment $ts_i$, $DS$ is expressed as: 
\begin{equation}
DS=\frac{\big|\;DR(ts_i) - \frac{1}{|N_{\epsilon}(ts_i)|}\sum\limits_{j}^{|N_{\epsilon}(ts_i)|}DR(ts_j)\;\big|}{\max\limits_{j} \big\{DR(ts_j)\big\} - \min\limits_{j} \big\{DR(ts_j)\big\}}
\end{equation}
$DS$ evaluates the difference between the core trajectory segment direction and the average direction of the clusters. A low $DS$ score indicates better homogeneity in term of direction.\newline

\noindent\textbf{Definition 14: The Homogeneity Score \textbf{(DS)}.}
$HS$ is the linear combination of the three aforementioned scores. 
It is expressed as follows:
\begin{equation}
HS = \omega_1\cdot FS + \omega_2\cdot SS + \omega_3\cdot DS
\label{hs}
\end{equation}
Where $\omega_1$, $\omega_2$ and $\omega_3$ are tunable weights to adjust the contribution of each score. The optimal cluster achieves the lowest $HS$.
\section{Experimental Results}
We conduct a set of experiments to show the effectiveness of our approach in terms of Sum of Squared Error and accuracy.  
\subsection{Experimental setup}
\begin{figure}[ht]
	\centering
	\includegraphics[scale=.4]{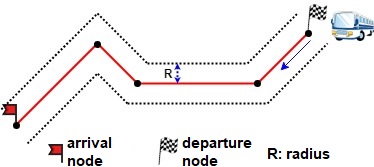}
	\caption{Full bus route from the departure station to the arrival station}
	\label{path}
\end{figure}

For our experiments, we use the public bus transport dataset of New York City\footnote{web.mta.info/developers/MTA-Bus-Time-historical-data.html}. The dataset tracks 90 bus services across New York city for a full day. The buses arrival and departure times are recorded along with the geolocation of each station. Each station has a unique id. Between two consecutive stations, we randomly generate 40 geolocations with unique ids to simulate trajectories of MCS.
\newline
The simulated sensors may be widespread and therefore can be out of the route segments of interest. To deal with this issue, we used the following heuristic to pre-filter  the irrelevant sensors. Given the historic journeys of vehicles, we establish the full regular route of every journey service such as a bus service. We consider the complete geolocations of every bus stop as illustrated in Fig. \ref{path} to derive the static service route. This data is static (e.g. station geolocations) and represent the journey service route as advertised by the service provider. Since our objective is to spatiotemporally  identify vehicles through MCS, we only consider the crowdsourced sensors within a buffer area with radius $R$ from the path line. Indeed, the objective is to filter out irrelevant MCS, i.e. the ones that do not contribute to identifying the journey vehicle and therefore the journey service.

To assess the effectiveness of the algorithms, we use the Sum of Squared Error (SSE). It reflects the overall compactness of the obtained data partition by calculating the distance between the center of each cluster and its associated objects. Therefore, the best clustering yields to the minimum SSE value which is computed as follows:
\begin{equation}
\begin{tabular}{c}
\( SSE = \sum_{i=1}^{num\_cluster} \bigg(\frac{1}{2|C_i|}  \sum\limits_{ts_k \in C_i}  \sum\limits_{ts_l \in C_i} d (ts_k,ts_l)    \bigg) \)\\
\end{tabular}
\end{equation}
We also propose a new index to evaluate the trajectory clustering results. Inspired by the Xie-Beni cluster validity index \cite{xie}, our proposed index named Tra-Xie-Beni (Tra-XB), takes into consideration the intra-cluster homogeneity as well as the inter-cluster separation. It is expressed as follows:
\begin{equation}
Tra\mbox{-}XB = \frac{\sum\limits_{i=1}^{num\_cluster} \sum\limits_{ts_j \in C_i}d(ts_j,cc_i)}{\min\limits_{i,j}d(cc_i,cc_j)}
\end{equation}
The nominator term calculates the distance between every cluster  $cc_i$ and $ts_j \in C_i$. This quantifies the compactness of every cluster. The denominator evaluates the separation between clusters which is the minimum distance between all cluster centers. Therefore, the best partition corresponds to the minimum value of Tra-XB.
In addition, we evaluate the spatial and temporal accuracy of both approaches by: 1- Calculating the spatial distance $d(n_a,\hat{n}_a)$ between the true destination node of the journey vehicle $n_a$ and the estimated destination point $\hat{n}_a$ computed by each algorithm.
2- Calculating the temporal error $att-\hat{att}$ between the actual average travel time $att$ of the journey vehicle and the average arrival time $\hat{att}$ computed by each algorithm.
The estimated parameters ($\hat{n_a}$, $\hat{att}$) are identified by determining the closest cluster center end point and its associated timestamp.
\newline
We provide a quantitative analysis for the first 30 timestamps of the schedule, although our findings can be reproduced for any desired period.
For the simulation settings, we set  $\epsilon = 0.002$  and $MinS = 17$ for the proposed algorithm while we choose for Traclus the optimal values of $MinS$ and $\epsilon$ i.e. the ones achieving the best performance. Similarly, we report the best performance achieved by ST-DBSCAN. We also set $\omega_1 = \omega_2 = \omega_3 =1$ and the preprocessing radius $R=10 m$.
\vspace{- 0.5 cm}
\begin{figure}[ht] 
	\subfigure[SSE index for the first 30 timestamps]{%
		\includegraphics[scale=.5]{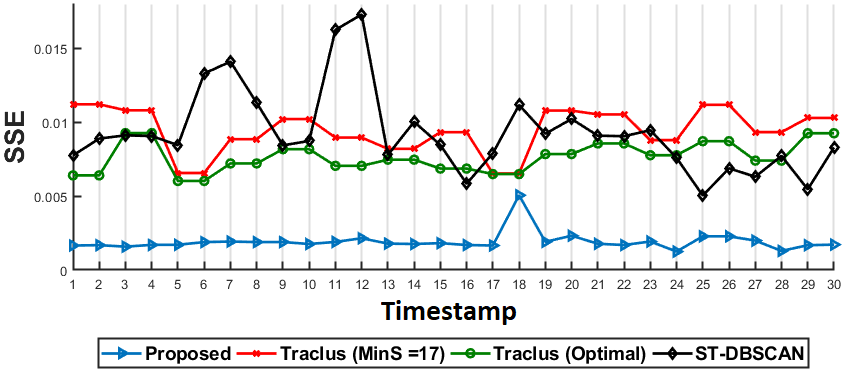} 
		\label{sse_30} 
	} 
	\subfigure[Tra-XB index for the next 30 timestamps]{%
		\includegraphics[scale=.5]{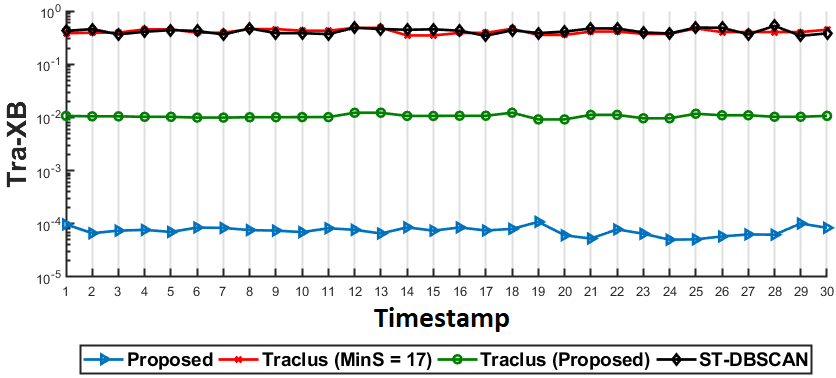} 
		\label{xie_30} 
	} 
	\caption{Spatiotemporal accuracy } 
\end{figure}
\vspace{- 0.5 cm}
\begin{figure}[ht] 
	\subfigure[Q64: spatial error]{%
		\includegraphics[scale=.5]{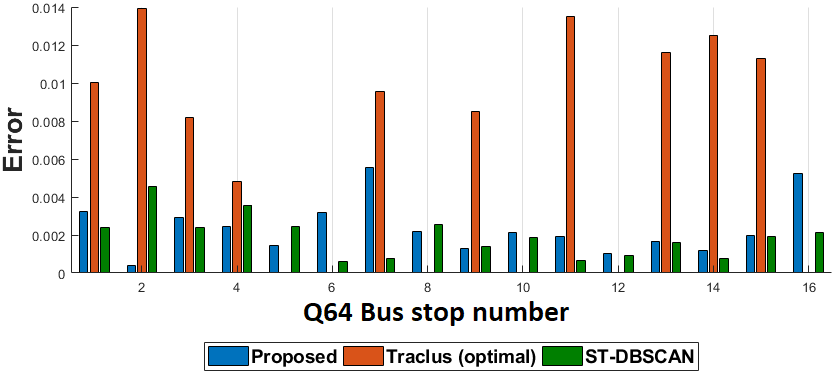} 
		\label{s_error} 
	} 
	\subfigure[Q64: temporal error]{%
		\includegraphics[scale=.5]{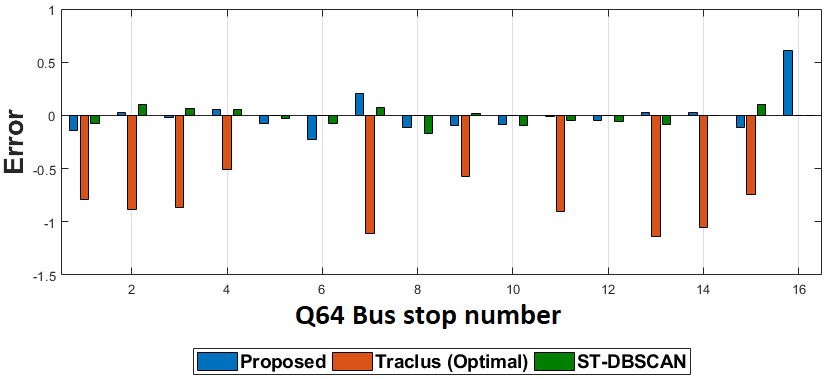} 
		\label{t_error} 
	} 
	\caption{Spatiotemporal error } 
\end{figure}
\vspace{- 0.5 cm}
\subsection{Discussion of Evaluation Results}
Fig. \ref{sse_30} depicts the SSE results of the algorithms for the first 30 timestamps of the schedule. We notice that the proposed clustering approach achieves better performance in terms of cluster compactness compared to ST-DBSCAN and Traclus with two settings:  Optimal $MinS$ value and $MinS=17$. This performance is confirmed by the Tra-XB index results illustrated in Fig. \ref{xie_30}. Therefore, we can conclude that the proposed algorithm achieves the best clustering performance with the best separation between clusters.\newline
Fig. \ref{s_error} and \ref{t_error} represent the spatiotemporal evaluation of Q64 bus service starting from 164 Street/Jewel Avenue station at midnight to 108 Street/Queens Boulevard arriving at 9 min past midnight. The whole journey consists of 16 route segments. First, we notice that Traclus algorithm was not able to identify a cluster (5$^{th}$ and 6$^{th}$ Q64 bus instance) and therefore failed to identify the journey vehicle. On the other hand, the proposed algorithm and ST-DBSCAN exhibit better spatiotemporal accuracy. For example, for the 11$^{th}$ Q64 bus instance, the spatial error achieved by the proposed algorithm is 0.002, 15 times less than the error achieved by Traclus. ST-DBSCAN achieves slightly better performance with spatial error = 0.0007.  Furthermore, the proposed approach and ST-DBSCAN compute $t_e$ with almost 100\% accuracy (error =-0.08 seconds and  -0.0475 seconds ) unlike Traclus which achieved a temporal error of around 1 second. \newline
We conclude that the proposed approach achieves better performance in accurately identifying vehicles and estimating their location and their arrival time to the next stop. It also achieves competitive performance compared to ST-DBSCAN. This performance can be explained by the capacity of the proposed approach to better capture the dynamism of the objects to be clustered since it takes into account several features such as speed, direction and following degree unlike density-based algorithm such as Traclus.
\section{Conclusion}
This paper proposed an approach to integrate real-time sensory data collected from multiple mobile crowdsourced sensors (MCS) to find a better journey service. We proposed and evaluated a clustering algorithm to select the right crowdsourced sensors, which enables the identification of the journey vehicles. This also helps to estimate journey services' location and  arrival time to the next stop. The proposed algorithm takes into account the following degree, speed and direction of the crowdsourced sensors to build clusters of moving objects. This cluster allows the identification of journey vehicles. Experimental results demonstrate the effectiveness of the proposed algorithm in achieving better performance in terms of cluster compactness compared to the existing approaches. 
In future work, we will analyze the computation complexity of our approach and develop an enhanced algorithm to detect clusters.  
Devising incentive mechanisms to encourage commuters to participate and contribute as a sensor is another interesting future work direction.
\section*{Acknowledgment}
This research was made possible by NPRP 9-224-1-049 grant from the Qatar National Research Fund (a member of The Qatar Foundation) \newline and DP160100149 and LE180100158 grants from Australian Research Council. The statements made herein are solely the responsibility of the authors.

\end{document}